\documentclass[11pt,a4paper]{article}
\usepackage{comment}
\usepackage[utf8]{inputenc}
\usepackage[T1]{fontenc}
\usepackage{lmodern}
\usepackage{amsmath,amssymb,amsthm,mathtools}
\usepackage{bm}
\usepackage{graphicx}
\usepackage{booktabs}
\usepackage{multirow}
\usepackage{enumitem}
\usepackage{algorithm}
\usepackage{algpseudocode}
\usepackage{authblk}
\usepackage[hidelinks]{hyperref}
\usepackage{geometry}
\usepackage{xcolor}
\title{Differential Analysis of Microbial Interaction Networks}

\author[1,*]{Marianna Milano}
\author[2,*]{Pietro Hiram Guzzi}
\affil[1]{Department of Experimental and Clinical Medicine, University of Catanzaro, Italy}
\affil[2]{Department of Surgical and Medical Sciences, Magna Graecia University of Catanzaro, 88100 Catanzaro, Italy}

\affil[*]{Corresponding author: \texttt{m.milano@unicz.it}}

\date{}

\begin{document}
\maketitle

\begin{abstract}
Microbiome studies increasingly indicate that disease-associated shifts cannot be understood from compositional changes alone. The functional architecture of microbial communities—encoded in patterns of association among microbial gene families—may reveal how these systems reorganize across biological conditions.

Here, we present a network-based framework for characterizing microbiome rewiring across conditions. The approach combines condition-specific network inference, differential network analysis and pathway enrichment to identify interactions that are gained, lost or altered between groups, with a specific focus on sex-dependent differences. We apply the framework to inflammatory bowel disease, type 2 diabetes and atherosclerotic cardiovascular disease, comparing male- and female-specific microbial gene-family networks within each disease context.

Across these settings, differential networks reveal extensive rewiring of microbial functional interactions, suggesting that microbiome alterations are shaped not only by changes in abundance but also by shifts in community organization. Importantly, pathway enrichment of rewired interactions uncovers functional signals that are not apparent from individual networks alone, highlighting latent disease- and sex-associated mechanisms. Code, data and supplementary information are available at \url{https://github.com/mmilano87/NetMicrobiome}
.
\end{abstract}

\section{Introduction}

The human microbiome is a complex and dynamic ecosystem composed of different microorganisms that coexist with the host and contribute to its physiological functioning. These microbial communities are involved in multiple biological processes, including metabolism, immune system regulation, and neurological activity, and are essential for nutrient processing, immune development, and cellular signaling \cite{fung2017interactions}. Therefore, the microbiota is widely recognized as a key component in maintaining host homeostasis \cite{lin2017role}.

Alterations in microbiome composition and organization have been associated with a broad range of pathological conditions, including metabolic, cardiovascular, and neurodegenerative diseases. In particular, changes in gut microbial composition have been linked to type 2 diabetes, inflammatory bowel disease, and Alzheimer’s disease, indicating that the microbiome may play an active role in disease onset and progression rather than acting as  marker \cite{manor2020health}. These effects are mediated by complex interaction patterns between host and microbial communities, where both commensal and pathogenic microorganisms modulate host signaling pathways, immune responses, and cellular processes \cite{pickard2017gut,gutierrez2022hospitalizations,agapito2013cloud4snp}.

Despite these advances, most microbiome studies rely on differential abundance or diversity measures, which primarily describe compositional variations across conditions and may only partially reflect the underlying interaction structure of microbial communities. Indeed, the microbiome can be naturally modeled as a complex system of interacting entities, where microbial taxa (i.e., microbial units such as species, genera, or operational taxonomic units) are linked through cooperative, competitive, and antagonistic relationships. In this context, network-based approaches provide a suitable framework to characterize microbial organization by representing taxa as nodes and their associations as edges, thus enabling a systems analysis of community structure \cite{srinivasan2024modeling}.

Within this framework, network analysis allows the identification of key structural properties, such as hubs and modules, which are often associated with functional relevance and system stability \cite{manor2020health}. Importantly, this representation also enables the comparison of network structures across conditions, providing a natural basis to investigate how microbial interactions change in different biological states.

A central concept emerging from this perspective is \textit{network rewiring}, defined as the reorganization of interaction patterns across conditions. In particular, rewiring captures variations in the presence, strength, or sign of associations between microbial entities, reflecting condition-specific modifications in network connectivity. By explicitly modeling these changes, rewiring analysis provides insights into the mechanisms underlying phenotypic variation that cannot be captured by composition-based approaches alone \cite{kim2012network}.

To systematically capture these changes, differential network analysis (DiNA) has been introduced as a powerful framework for comparing networks across different conditions. Within the context of network rewiring, DiNA provides a formal approach to quantify how interaction patterns are reorganized between distinct biological states. In particular, given two networks representing different conditions, such as healthy versus diseased or male versus female subjects, DiNA aims to identify variations in connectivity patterns by explicitly comparing their edge structures \cite{guzzi2024non}.

By focusing on differences at the edge level, rather than solely on node changes, differential networks enable the identification of interactions that are gained, lost, or significantly altered between conditions. These changes directly reflect rewiring events, capturing modifications in the presence, strength, or sign of associations between entities. As a result, DNA provides a more comprehensive characterization of system alterations, revealing condition-specific interaction patterns.
Overall, differential network analysis represents a key tool to investigate how biological systems reorganize under different conditions, allowing the identification of interaction changes that underlie phenotypic variability.

In this study, we propose a network-based framework to investigate microbiome alterations across different biological conditions, with a particular focus on sex-specific differences. Our approach integrates network construction, rewiring analysis, and differential network modeling to characterize changes in microbial interaction patterns.

Specifically, we construct condition-specific microbiome networks separately for male and female subjects across three disease i.e, inflammatory (IBD), metabolic (T2D), and cardiovascular (ACVD) conditions. Differential networks are then derived to identify significant variations in associations between microbial gene families, capturing interaction changes between the two groups. 

To further interpret these alterations from a functional perspective, we perform pathway enrichment analysis on the gene families involved in rewiring, allowing us to link network changes to biological processes.

Our results show that differential network analysis captures meaningful rewiring patterns that are not detectable through standard abundance-based approaches. In particular, we identify sex-specific interaction changes and novel pathways that emerge exclusively from differential networks, highlighting the role of interaction variations in shaping microbiome-associated phenotypes. Overall, these findings support the adoption of a systems perspective to better understand the mechanisms underlying microbiome-related diseases.

\section{Motivation}

Network-based approaches have become a powerful framework for investigating the structure and functionality of microbial communities, enabling the characterization of interactions among microbial entities beyond simple abundance profiles \cite{faust2012microbial, weiss2016correlation}. However, several methodological challenges still limit their effectiveness in capturing condition-specific alterations in microbiome organization.

A major limitation of existing studies lies in the adoption of global feature sets shared across conditions, which may obscure interaction patterns that are specific to a given biological state. In this setting, networks inferred from different groups are often directly compared without accounting for the fact that relevant features may differ across conditions.

In addition, most microbiome studies rely on differential abundance analysis to identify condition-associated changes \cite{macklaim2013comparative, hou2022microbiota}. While effective at detecting variations in microbial composition, these approaches do not capture how relationships among microbial entities are reorganized. As a result, alterations occurring at the interaction level may remain undetected.

The intrinsic properties of microbiome data further complicate network inference. Microbial abundance data are characterized by sparsity, compositional constraints, and non-linear dependencies, which can introduce biases and instability in correlation-based methods \cite{gloor2017microbiome, kurtz2015sparse}. These issues may affect the reliability of inferred networks and limit the identification of robust interaction patterns.

Furthermore, biological stratification factors, such as sex-specific differences, are often not explicitly incorporated into network-based analyses, despite increasing evidence of their relevance in microbiome-associated diseases \cite{markle2013sex, ahmed2021sex,javsarevic2016sex}. Accounting for such factors is essential for a more accurate characterization of microbiome organization.

Motivated by these limitations, we propose a framework aimed at modeling microbiome alterations at the level of interaction dynamics. In particular, our approach is designed to (i) infer condition-specific networks based on group-dependent feature selection, (ii) identify rewiring patterns through differential network analysis, and (iii) provide a functional interpretation of these changes via pathway enrichment analysis.

By focusing on interaction variations and integrating biological stratification, the proposed framework enables a more comprehensive representation of microbiome organization across different conditions.

\section{Related Work}

Network-based approaches have become a widely adopted framework for modeling the human microbiome, as they enable the characterization of interactions among microbial entities beyond simple abundance profiles \cite{guven2017structural}. By representing microbial taxa or gene families as nodes and their associations as connections, these methods provide a system view of community organization and have been extensively used to investigate microbiome structure and its relationship with disease.

To address the specific challenges of microbiome data, such as compositionality and sparsity, several computational methods have been developed for network inference. SparCC (Sparse Correlations for Compositional data) \cite{friedman2012inferring} estimates correlations between microbial taxa by inferring associations between latent absolute abundances through log-ratio transformations, under the assumption of a sparse underlying correlation structure . 

CoNet \cite{faust2012microbial} adopts an ensemble strategy by combining multiple association measures, including correlation, dissimilarity, and information-theoretic metrics, and identifies statistically significant interactions through permutation and bootstrap procedures . 

SPIEC-EASI (Sparse Inverse Covariance Estimation for Ecological Association Inference)\cite{kurtz2015sparse} relies on graphical models and sparse inverse covariance estimation to infer direct associations, reducing the impact of indirect correlations and compositional effects . 

NetCoMi \cite{peschel2021netcomi} provides a comprehensive framework for microbiome network construction, comparison, and analysis, integrating different inference methods and supporting the computation of network properties for downstream analysis .

While these methods have significantly advanced microbiome network inference, they are primarily focused on the construction and analysis of individual networks. Most existing studies rely on topological descriptors or differential abundance at the node level, without explicitly addressing how interaction patterns change across conditions. As a result, the variability and reorganization of microbial interactions remain only partially explored.

In parallel, differential network analysis has been extensively investigated in other domains, particularly in gene co-expression studies, where it is used to identify changes in connectivity patterns between biological states \cite{guzzi2024non, kim2012network}. These approaches have demonstrated that variations in network structure can reveal system alterations that are not detectable through feature-based analyses alone.

Building on this paradigm, recent studies have started to explore the application of differential network analysis in the microbiome domain. Early approaches rely on correlation-based network construction, where microbial association networks are inferred independently for each condition and compared through global or topological properties \cite{espinoza2022differential}. While these methods highlight the potential of network-based comparisons to uncover condition-specific interaction patterns, they are limited by their descriptive nature and do not explicitly model differences between networks.

More recent methods aim to incorporate statistical modeling into differential microbiome network analysis. For example, Ahn et al. \cite{ahn2024differential} proposed SOHPIE-DNA, a regression-based framework that detects differential connectivity by modeling node centrality measures as a function of covariates. Similarly, tools such as NetCoMi \cite{peschel2021netcomi} and MDiNE \cite{mcgregor2020mdine} enable the construction and comparison of microbial networks across conditions, focusing on differences in network topology or connectivity patterns. More advanced probabilistic approaches, such as the Bayesian covariance regression model TRECOR \cite{xu2026bayesian}, allow the estimation of networks whose covariance structure varies with covariates, capturing context-dependent changes in microbial interactions.

Despite these advances, existing approaches differ substantially in their objectives and modeling assumptions. In particular, most methods focus on node connectivity or global network properties, or model covariance structures without explicitly characterizing interaction changes. As a consequence, they do not provide a unified and formal framework to define and systematically quantify network rewiring across conditions.

For this reason, a direct comparison with existing tools is not straightforward, as they address related but fundamentally different problems. Network inference methods such as SparCC, CoNet, SPIEC-EASI, and NetCoMi are primarily designed for constructing microbial association networks, while differential approaches in both microbiome and other domains focus on connectivity or covariance changes rather than explicitly modeling variations in interaction patterns.

In this study, we propose a unified framework that integrates condition-specific network construction and differential network analysis to characterize changes in interaction patterns across multiple conditions and sexes. By introducing a formal definition of rewiring and enabling a systematic comparison of networks, our approach provides a coherent and interpretable framework to study microbiome organization.

To the best of our knowledge, this work represents one of the first attempts to systematically investigate microbiome rewiring through differential networks in a multi-condition and sex-aware setting, addressing a key gap in current methodologies.

\section{Methodology}

The analysis of the gut microbiome is inherently complex due to the high dimensionality and interdependence of microbial functional components. For this reason, in this study we adopt a network-based approach to model the microbiome as a system of interacting gene families, enabling the characterization of coordinated functional relationships. This representation enables the investigation of coordinated patterns of microbial activity and provides a natural framework to study how such interactions vary across biological conditions.

We investigate microbiome organization by explicitly considering both disease status (healthy vs diseased) and sex (male vs female). Our objective is to characterize how functional relationships among microbial gene families vary across these groups, with the goal to capture differences in their interaction structure.

To this end, we adopt an interaction perspective, where microbiome alterations are captured through variations in the structure and strength of relationships between gene families inferred from group-specific data. This approach allows to take in to account  feature level analysis and to directly model differences in the organization of microbial functional networks.

Our methodological pipeline consists of three main steps. First, we construct group-specific functional networks independently for each condition and sex, using a dynamic feature selection strategy that allows each network to reflect the specific functional composition of the corresponding group. Second, we compare these networks to identify changes in microbial interactions between groups, capturing variations in both the presence and strength of associations. Finally, we apply pathway enrichment analysis on the set of genes involved in the identified interaction changes, in order to  highlight the biological processes most affected by the investigated factors.

\begin{figure}[H]
\centering
\includegraphics[ width=6in]{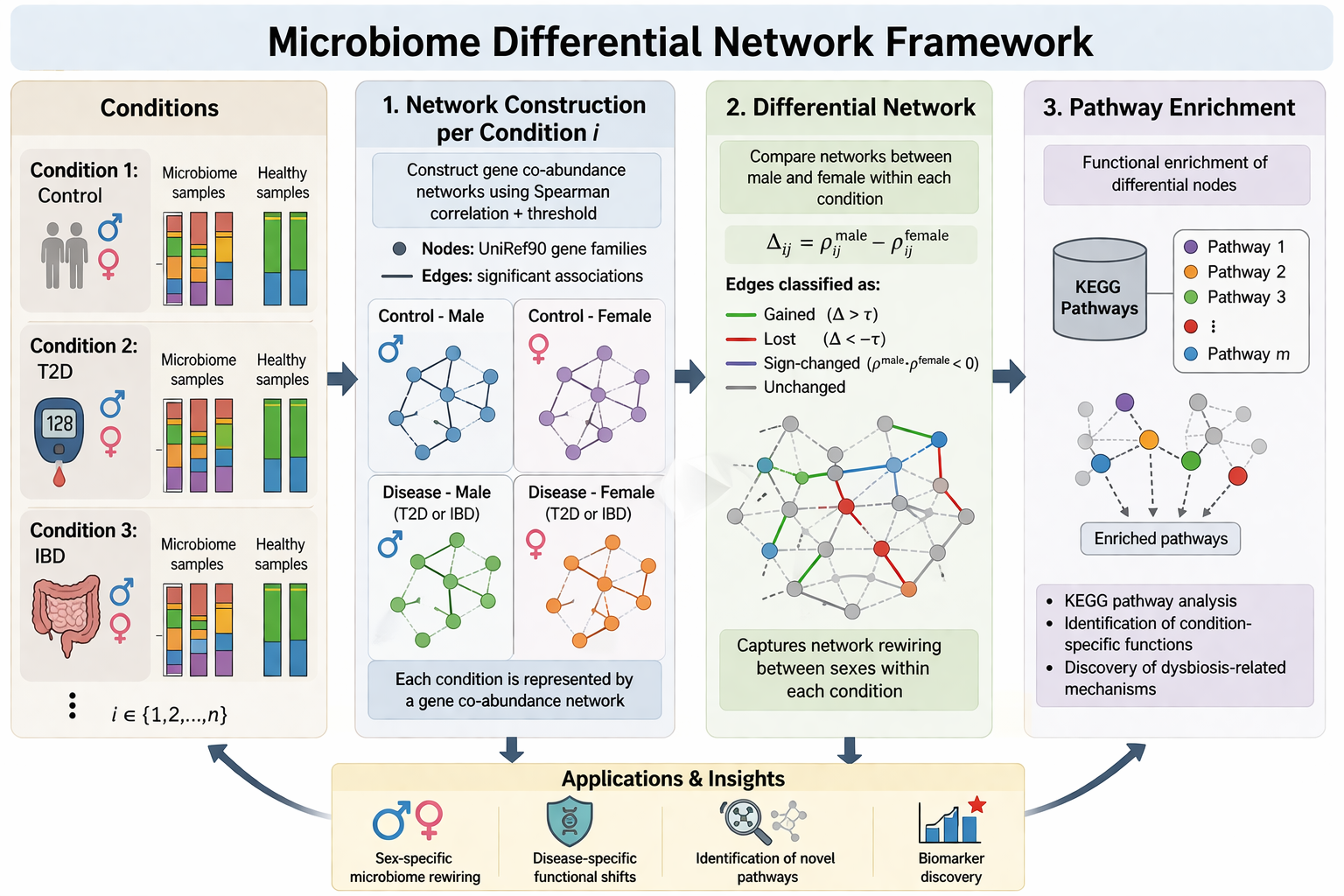}
\caption{
\textbf{Microbiome differential network analysis framework.}
Microbiome data from multiple conditions (for example,control, T2D, and IBD) are used to construct sex-specific gene co-abundance networks. Differential networks are computed as $\Delta_{ij} = \rho_{ij}^{\text{male}} - \rho_{ij}^{\text{female}}$, with edges classified as gained, lost, or sign-changed, capturing network rewiring. Functional enrichment of differential nodes identifies condition-specific KEGG pathways and highlights sex-specific microbiome alterations and potential biomarkers.
}

\label{fig:wf}   
\end{figure}

The overall workflow is illustrated in Figure  \ref{fig:wf}.

\subsection{Construction of Functional Networks}

To study the gut microbiome within a network-based framework, we represent each microbial community as an undirected weighted network, where nodes correspond to microbial gene families  and edges represent statistical associations between their abundance profiles across samples.

The analysis starts from the gene family abundance matrix, where rows correspond to gene families and columns to samples. For a given group $g$, let $X^{(g)}$ denote the matrix containing the abundances of all gene families across the samples belonging to that group. Each entry $X^{(g)}_{ij}$ represents the abundance of gene family $i$ in sample $j$. In this setting, each gene family is treated as a feature describing the functional composition of the microbiome.

Since microbiome data are typically sparse and high-dimensional, we applied a group-specific filtering strategy to retain only informative features. In particular, feature selection was performed independently within each group, so that each network is constructed from the subset of gene families that are relevant for that specific condition. As a result, the set of nodes is not fixed in advance but is determined by the data, allowing each network to better reflect the underlying biological variability.

The first filtering step is based on prevalence, defined as the fraction of samples in which a given gene family is observed:

\begin{equation}
\mathrm{Prev}_i^{(g)} = \frac{1}{n_g} \sum_{j=1}^{n_g} \mathbb{I}(X^{(g)}_{ij} > 0)
\end{equation}

where $\mathbb{I}(X^{(g)}_{ij} > 0)$ is equal to 1 if the gene family is present in sample $j$, and 0 otherwise. In other words, prevalence measures how frequently a gene family appears across samples.

We retained only gene families with prevalence between $0.1$ and $0.9$. The lower bound removes very rare features that appear in only a few samples and may lead to unstable or spurious correlations. The upper bound removes highly ubiquitous features that are present in almost all samples and therefore provide little discriminatory information across conditions.

A second filtering step was based on mean abundance. For each gene family, we computed

\begin{equation}
\mu_i^{(g)} = \frac{1}{n_g} \sum_{j=1}^{n_g} X^{(g)}_{ij}
\end{equation}

and retained only those features satisfying $\mu_i^{(g)} \geq 10^{-6}$. This threshold was introduced to remove gene families that, although present in multiple samples, exhibit extremely low abundance values. In microbiome data, such low-abundance features are often dominated by measurement noise and may lead to unstable correlation estimates. Therefore, this filtering step ensures that only features contributing a meaningful signal to the association structure are retained.

We then quantified feature variability within each group through the variance

\begin{equation}
\left(\sigma_i^{2}\right)^{(g)} = \frac{1}{n_g} \sum_{j=1}^{n_g} \left(X^{(g)}_{ij} - \mu_i^{(g)}\right)^2
\end{equation}

and retained only features with variance greater than or equal to $10^{-8}$. This threshold allows us to exclude gene families with nearly constant abundance across samples, which do not contribute to the identification of meaningful associations. In practice, very low-variance features tend to generate artificially inflated or non-informative correlations.

These thresholds were chosen to balance noise reduction and information preservation, and are consistent with common preprocessing strategies adopted in microbiome network analysis \cite{faust2012microbial, peschel2021netcomi}. Unlike fixed top-$k$ selection strategies, this approach preserves all features that satisfy minimal signal and variability requirements, remaining consistent with the dynamic nature of the node set.

After filtering, the abundance values were log-transformed to stabilize the distribution and reduce the impact of highly skewed abundances. Specifically, we defined

\begin{equation}
\widetilde{X}^{(g)}_{ij} = \log\left(X^{(g)}_{ij} + \epsilon\right),
\end{equation}

where $\epsilon = 10^{-6}$ is a pseudocount introduced to avoid undefined values for zero entries.

Pairwise associations between the selected gene families were estimated using the Spearman correlation coefficient, which is well suited for microbiome data as it does not assume normality and is robust to monotonic, potentially non-linear relationships \cite{faust2012microbial}. 

For each pair of features $(i,k)$ in group $g$, we computed

\begin{equation}
\rho_{ik}^{(g)} = \mathrm{corr}_S\left(\widetilde{X}^{(g)}_{i\cdot}, \widetilde{X}^{(g)}_{k\cdot}\right),
\end{equation}

where $\widetilde{X}^{(g)}_{i\cdot}$ and $\widetilde{X}^{(g)}_{k\cdot}$ represent the abundance profiles of the two gene families across all samples in group $g$. These correlation values quantify how similarly two gene families vary across samples and are therefore used to define potential functional associations.

An edge between two nodes $i$ and $k$ was introduced if the absolute value of their correlation exceeded a threshold $\tau = 0.3$, that is,

\begin{equation}
\left| \rho_{ik}^{(g)} \right| \geq \tau.
\end{equation}

In this way, nodes correspond to the gene families retained after filtering, while edges represent statistically relevant co-abundance relationships between them.

The resulting network for group $g$ is defined as

\begin{equation}
G^{(g)} = \left(V^{(g)}, E^{(g)}\right),
\end{equation}

where $V^{(g)}$ is the set of selected gene families and

\begin{equation}
E^{(g)} = \left\{ (i,k) : \left| \rho_{ik}^{(g)} \right| \geq \tau \right\}.
\end{equation}

Each edge is assigned a weight equal to $\rho_{ik}^{(g)}$, thus preserving both the strength and the sign of the association. Since feature selection is performed independently for each group, the resulting networks may differ both in their node sets and in their connectivity patterns. This reflects the fact that different conditions may be characterized by distinct functional profiles and interaction structures.

To summarize the overall organization of each network, we computed standard topological descriptors, including the number of nodes and edges, density, number of connected components, size of the largest connected component, average clustering coefficient, and average betweenness centrality. These measures provide a global characterization of network structure and are used to compare different conditions before performing the differential network analysis.

\subsection{Differential Network Construction }

To investigate sex-specific differences in the functional organization of the gut microbiome, we performed a differential network analysis aimed at identifying changes in the interaction structure between microbial gene families across conditions.

Within a network-based framework, biological differences are not only reflected by variations in the abundance of individual features, but also by changes in the relationships among them. These changes, referred to as \textit{network rewiring}, describe how interactions between components of the system are reorganized across conditions. Capturing such rewiring patterns enables the identification of functional alterations that cannot be detected through feature analysis alone.

Given two functional networks inferred from different conditions, denoted as G$_{1}$ = (V$_{1}$, E$_{1}$) and  G$_{2}$ = (V$_{2}$, E$_{2}$), differential network analysis aims to quantify variations in their connectivity patterns. Each edge is associated with a weight corresponding to the Spearman correlation between gene family abundance profiles.

Because the two networks are constructed independently using a dynamic feature selection strategy, they may differ in both node sets and edge sets. For this reason, the comparison is performed on the union of all observed edges. Let $\mathcal{E}$ = E $_{1}$ $\cup$ E$_{2}$ denote the set of all edges appearing in at least one of the two networks.

For each edge $(i,k) \in \mathcal{E}$, we define its association values in the two networks as:

\begin{equation}
r^{(1)}_{ik} =
\begin{cases}
\rho^{(1)}_{ik}, & \text{if } (i,k) \in E_{1},\\
0, & \text{otherwise,}
\end{cases}
\end{equation}

\begin{equation}
r^{(2)}_{ik} =
\begin{cases}
\rho^{(2)}_{ik}, & \text{if } (i,k) \in E_{2},\\
0, & \text{otherwise.}
\end{cases}
\end{equation}

The differential signal associated with each edge is defined as:

\begin{equation}
\Delta_{ik} = r^{(1)}_{ik} - r^{(2)}_{ik}.
\end{equation}

To identify meaningful rewiring events, we introduce two thresholds. The first is an association threshold $\tau = 0.3$, used to determine whether an interaction is considered present in a given network. This threshold was selected to retain moderate-to-strong correlations while filtering out weak associations that are more likely to arise from noise, which is a common issue in high-dimensional and sparse microbiome data. Values around 0.3 are widely adopted in microbiome network analysis as a compromise between sensitivity and robustness, allowing the identification of biologically meaningful interactions without excessively increasing network density \cite{faust2012microbial, peschel2021netcomi}.

The second is a differential threshold $\delta = 0.3$, which defines the minimum change in association strength required to identify a significant variation between the two networks. By imposing this threshold, we ensure that only substantial differences in interaction strength are interpreted as rewiring events, avoiding the inclusion of minor fluctuations that may not be biologically relevant.

Based on these criteria, each edge is classified into one of the following categories.

An edge is classified as \textit{sign-changed} if it is present in both networks and its sign is reversed, i.e.,

\begin{equation}
|r^{(1)}_{ik}| \geq \tau, \quad |r^{(2)}_{ik}| \geq \tau, \quad \text{and} \quad r^{(1)}_{ik} \cdot r^{(2)}_{ik} < 0.
\end{equation}

An edge is classified as \textit{exclusive to $G_{1}$} if it is present in $G_{1}$ but absent in $G_{2}$, that is,

\begin{equation}
|r^{(1)}_{ik}| \geq \tau \quad \text{and} \quad |r^{(2)}_{ik}| < \tau.
\end{equation}

Conversely, an edge is classified as \textit{exclusive to $G_{2}$} if it is present in $G_{2}$ but absent in $G_{1}$, i.e.,

\begin{equation}
|r^{(1)}_{ik}| < \tau \quad \text{and} \quad |r^{(2)}_{ik}| \geq \tau.
\end{equation}

Finally, an edge is classified as \textit{differentially weighted} if it is present in at least one of the two networks and exhibits a substantial difference in strength:

\begin{equation}
\max\left(|r^{(1)}_{ik}|, |r^{(2)}_{ik}|\right) \geq \tau \quad \text{and} \quad |\Delta_{ik}| \geq \delta.
\end{equation}

The differential network is therefore defined as

\begin{equation}
G_{\Delta} = (V_{\Delta}, E_{\Delta}),
\end{equation}

where $E_{\Delta}$ contains all edges classified as rewired interactions. Each edge is assigned a weight equal to the magnitude of the differential signal:

\begin{equation}
w^{\Delta}_{ik} = |\Delta_{ik}|.
\end{equation}

To quantify rewiring at the node level, we define a rewiring score for each node as:

\begin{equation}
R_i = \sum_{k : (i,k) \in E_{\Delta}} |\Delta_{ik}|.
\end{equation}

Nodes with high rewiring scores are those whose connectivity patterns change most strongly between the two networks.

The different types of edge rewiring identified by our framework are depicted in Figure \ref{fig:dn}.
\begin{figure}[H] 
\centering
\includegraphics[ width=4in]{img/DN.png} 
\caption{ Examples of edge rewiring between two functional networks $G_{1}$ and $G_{2}$. Nodes represent microbial gene families, and edges represent associations between them. The figure illustrates four types of rewiring: (a) sign change, where the interaction reverses direction; (b) interaction present only in $G_{1}$; (c) interaction present only in $G_{2}$; and (d) interaction present in both networks with different strength. Edges are defined using a correlation threshold $\tau$, and differences are evaluated using a differential threshold $\delta$.} 
\label{fig:dn} 
\end{figure}

\subsection{Pathway Enrichment Analysis}

To investigate the biological relevance of the identified network modules, we performed a pathway enrichment analysis aimed at linking gene interactions to higher functional processes.

Starting from the constructed networks, we first selected representative subsets of nodes for each condition. In particular, we considered the most relevant nodes in male and female networks based on their connectivity, as well as the top rewired nodes extracted from the differential network. This strategy allows us to focus on the most informative components of each network, capturing both condition-specific structure and rewiring effects.

For each selected node set, we computed a module activity profile by aggregating the abundance of the corresponding gene families across samples. This profile was then compared with pathway abundance data derived from the same cohort, enabling a direct association between network modules and biological pathways.

The association between each module and pathway was assessed using correlation analysis, allowing us to identify pathways whose variation is strongly linked to the activity of the considered network modules. Only statistically significant associations were retained after multiple testing correction, ensuring robustness of the results.

This approach differs from classical enrichment methods based solely on gene lists, as it integrates both network topology and functional activity. In particular, by leveraging the differential network, we are able to identify pathways specifically associated with rewiring processes rather than static network structures.

Finally, we compared the pathways identified in male, female, and differential networks. This comparison allowed us to highlight pathways that are shared across conditions as well as those uniquely emerging from the differential network. These latter pathways represent novel functional signatures that are not detectable when analyzing the individual networks separately, thus providing additional insight into the biological mechanisms underlying condition-specific differences.

Overall, this analysis enables a comprehensive interpretation of network rewiring in terms of biological functions, bridging the gap between topological changes and pathway dynamics.

\section{Conclusions}

In this work, we introduced a network-based framework to investigate microbiome alterations by modeling differences in associations between microbial gene families across conditions. The proposed approach integrates condition-specific network construction, differential network analysis, and pathway enrichment to capture rewiring patterns between male and female subjects.

Microbiome data were first processed to construct co-abundance networks for each condition and sex. Subsequently, differential networks were derived by comparing male and female networks, allowing the identification of interactions that are gained, lost, or altered between the two groups. This procedure enabled the  characterization of rewiring patterns across three disease contexts,  inflammatory bowel disease (IBD), type 2 diabetes (T2D), and atherosclerotic cardiovascular disease (ACVD).

The analysis revealed that microbiome alterations are largely driven by changes in the structure of associations between gene families. In particular, control conditions exhibited a higher number of rewired interactions, suggesting a more flexible and heterogeneous organization, whereas disease conditions showed a reduction in interaction variability, indicating a more constrained network structure.

Differential networks were used to characterize condition-specific rewiring by identifying changes in associations between microbial gene families. The functional interpretation of these rewiring events was subsequently performed through pathway enrichment analysis. Enrichment analysis highlighted metabolic and biosynthetic pathways, including amino acid metabolism, carbohydrate degradation, and energy-related processes, suggesting their involvement in disease-associated microbiome alterations. Importantly, several pathways were detected exclusively in differential networks, indicating that these functional signals emerge only when considering differences in interactions rather than individual networks.

Overall, the proposed framework provides a comprehensive approach to study microbiome organization across conditions, emphasizing the importance of considering interaction changes between microbial gene families. This perspective enables the identification of condition-specific mechanisms and supports a more detailed interpretation of microbiome-associated alterations.

From a translational perspective, this approach may contribute to improving the characterization of disease-associated microbiome profiles and supports future applications in biomarker discovery and personalized medicine. Further developments may include the integration of additional data layers, such as taxonomic profiles and host-related information, to better understand microbiome-host interactions and their role in complex diseases.

\section{Data availability}
Code, Data and Supplementary Information are available at \url{https://github.com/mmilano87/NetMicrobiome}.

 \section{Acknowledgments}
This work has been partially supported by the OFIDIAPlus (Operational Fire Danger preventIon plAtform Plus) project under the Interreg VI-A Greece-Italy 2021-2027 programme.\\
We acknowledge the support of the "HeartTrack: Driving the Patient Journey in Heart Failure” project funded by PR CALABRIA FESR FSE 2021 – 2027.

 \bibliographystyle{elsarticle-num} 
  \bibliography{biblio}

@article{gutierrez2022hospitalizations,
  title={Hospitalizations associated with mental health conditions among adolescents in the US and France during the COVID-19 pandemic},
  author={Guti{\'e}rrez-Sacrist{\'a}n, Alba and Serret-Larmande, Arnaud and Hutch, Meghan R and S{\'a}ez, Carlos and Aronow, Bruce J and Bhatnagar, Surbhi and Bonzel, Clara-Lea and Cai, Tianxi and Devkota, Batsal and Hanauer, David A and others},
  journal={JAMA network open},
  volume={5},
  number={12},
  pages={e2246548},
  year={2022}
}

@article{lin2017role,
  title={Role of intestinal microbiota and metabolites on gut homeostasis and human diseases},
  author={Lin, Lan and Zhang, Jianqiong},
  journal={BMC immunology},
  volume={18},
  number={1},
  pages={2},
  year={2017},
  publisher={Springer}
}

@article{guzzi2024non,
  title={Non parametric differential network analysis: a tool for unveiling specific molecular signatures},
  author={Guzzi, Pietro Hiram and Roy, Arkaprava and Milano, Marianna and Veltri, Pierangelo},
  journal={BMC bioinformatics},
  volume={25},
  number={1},
  pages={359},
  year={2024},
  publisher={Springer}
}

@article{srinivasan2024modeling,
  title={Modeling microbial community networks: methods and tools for studying microbial interactions},
  author={Srinivasan, Shanchana and Jnana, Apoorva and Murali, Thokur Sreepathy},
  journal={Microbial ecology},
  volume={87},
  number={1},
  pages={56},
  year={2024},
  publisher={Springer}
}

@article{manor2020health,
  title={Health and disease markers correlate with gut microbiome composition across thousands of people},
  author={Manor, Ohad and Dai, Chengzhen L and Kornilov, Sergey A and Smith, Brett and Price, Nathan D and Lovejoy, Jennifer C and Gibbons, Sean M and Magis, Andrew T},
  journal={Nature communications},
  volume={11},
  number={1},
  pages={5206},
  year={2020},
  publisher={Nature Publishing Group UK London}
}

@article{pickard2017gut,
  title={Gut microbiota: Role in pathogen colonization, immune responses, and inflammatory disease},
  author={Pickard, Joseph M and Zeng, Melody Y and Caruso, Roberta and N{\'u}{\~n}ez, Gabriel},
  journal={Immunological reviews},
  volume={279},
  number={1},
  pages={70--89},
  year={2017},
  publisher={Wiley Online Library}
}

@article{weiss2016correlation,
  title={Correlation detection strategies in microbial data sets vary widely in sensitivity and precision},
  author={Weiss, Sophie and Van Treuren, Will and Lozupone, Catherine and Faust, Karoline and Friedman, Jonathan and Deng, Ye and Xia, Li Charlie and Xu, Zhenjiang Zech and Ursell, Luke and Alm, Eric J and others},
  journal={The ISME journal},
  volume={10},
  number={7},
  pages={1669--1681},
  year={2016},
  publisher={Oxford University Press}
}

@article{friedman2012inferring,
  title={Inferring correlation networks from genomic survey data},
  author={Friedman, Jonathan and Alm, Eric J},
  year={2012},
  publisher={Public Library of Science San Francisco, USA}
}

@article{guven2017structural,
  title={Structural host-microbiota interaction networks},
  author={Guven-Maiorov, Emine and Tsai, Chung-Jung and Nussinov, Ruth},
  journal={PLoS computational biology},
  volume={13},
  number={10},
  pages={e1005579},
  year={2017},
  publisher={Public Library of Science San Francisco, CA USA}
}

@article{hou2022microbiota,
  title={Microbiota in health and diseases},
  author={Hou, Kaijian and Wu, Zhuo-Xun and Chen, Xuan-Yu and Wang, Jing-Quan and Zhang, Dongya and Xiao, Chuanxing and Zhu, Dan and Koya, Jagadish B and Wei, Liuya and Li, Jilin and others},
  journal={Signal transduction and targeted therapy},
  volume={7},
  number={1},
  pages={135},
  year={2022},
  publisher={Nature Publishing Group UK London}
}

@article{javsarevic2016sex,
  title={Sex differences in the gut microbiome--brain axis across the lifespan},
  author={Ja{\v{s}}arevi{\'c}, Eldin and Morrison, Kathleen E and Bale, Tracy L},
  journal={Philosophical Transactions of the Royal Society B: Biological Sciences},
  volume={371},
  number={1688},
  year={2016},
  publisher={The Royal Society}
}

@article{ahmed2021sex,
  title={Sex differences in the intestinal microbiome: interactions with risk factors for atherosclerosis and cardiovascular disease},
  author={Ahmed, Shamon and Spence, J David},
  journal={Biology of sex Differences},
  volume={12},
  number={1},
  pages={35},
  year={2021},
  publisher={Springer}
}

@article{markle2013sex,
  title={Sex differences in the gut microbiome drive hormone-dependent regulation of autoimmunity},
  author={Markle, Janet GM and Frank, Daniel N and Mortin-Toth, Steven and Robertson, Charles E and Feazel, Leah M and Rolle-Kampczyk, Ulrike and Von Bergen, Martin and McCoy, Kathy D and Macpherson, Andrew J and Danska, Jayne S},
  journal={Science},
  volume={339},
  number={6123},
  pages={1084--1088},
  year={2013},
  publisher={American Association for the Advancement of Science}
}

@article{gloor2017microbiome,
  title={Microbiome datasets are compositional: and this is not optional},
  author={Gloor, Gregory B and Macklaim, Jean M and Pawlowsky-Glahn, Vera and Egozcue, Juan J},
  journal={Frontiers in microbiology},
  volume={8},
  pages={2224},
  year={2017},
  publisher={Frontiers Media SA}
}

@article{macklaim2013comparative,
  title={Comparative meta-RNA-seq of the vaginal microbiota and differential expression by Lactobacillus iners in health and dysbiosis},
  author={Macklaim, Jean M and Fernandes, Andrew D and Di Bella, Julia M and Hammond, Jo-Anne and Reid, Gregor and Gloor, Gregory B},
  journal={Microbiome},
  volume={1},
  number={1},
  pages={12},
  year={2013},
  publisher={Springer}
}

@article{kurtz2015sparse,
  title={Sparse and compositionally robust inference of microbial ecological networks},
  author={Kurtz, Zachary D and M{\"u}ller, Christian L and Miraldi, Emily R and Littman, Dan R and Blaser, Martin J and Bonneau, Richard A},
  journal={PLoS computational biology},
  volume={11},
  number={5},
  pages={e1004226},
  year={2015},
  publisher={Public Library of Science San Francisco, CA USA}
}

@article{fung2017interactions,
  title={Interactions between the microbiota, immune and nervous systems in health and disease},
  author={Fung, Thomas C and Olson, Christine A and Hsiao, Elaine Y},
  journal={Nature neuroscience},
  volume={20},
  number={2},
  pages={145--155},
  year={2017},
  publisher={Nature Publishing Group US New York}
}

@article{faust2012microbial,
  title={Microbial co-occurrence relationships in the human microbiome},
  author={Faust, Karoline and Sathirapongsasuti, J Fah and Izard, Jacques and Segata, Nicola and Gevers, Dirk and Raes, Jeroen and Huttenhower, Curtis},
  journal={PLoS computational biology},
  volume={8},
  number={7},
  pages={e1002606},
  year={2012},
  publisher={Public Library of Science San Francisco, USA}
}

@article{peschel2021netcomi,
  title={NetCoMi: network construction and comparison for microbiome data in R},
  author={Peschel, Stefanie and M{\"u}ller, Christian L and Von Mutius, Erika and Boulesteix, Anne-Laure and Depner, Martin},
  journal={Briefings in bioinformatics},
  volume={22},
  number={4},
  pages={bbaa290},
  year={2021},
  publisher={Oxford University Press}
}

@article{kim2012network,
  title={Network rewiring is an important mechanism of gene essentiality change},
  author={Kim, Jinho and Kim, Inhae and Han, Seong Kyu and Bowie, James U and Kim, Sanguk},
  journal={Scientific reports},
  volume={2},
  number={1},
  pages={900},
  year={2012},
  publisher={Nature Publishing Group UK London}
}

@article{ahn2024differential,
  title={Differential network connectivity analysis for microbiome data adjusted for clinical covariates using jackknife pseudo-values},
  author={Ahn, Seungjun and Datta, Somnath},
  journal={BMC bioinformatics},
  volume={25},
  number={1},
  pages={117},
  year={2024},
  publisher={Springer}
}

@article{xu2026bayesian,
  title={Bayesian covariance regression for differential network analysis of zero-inflated microbiome data},
  author={Xu, Zichun and Ma, Jing},
  journal={arXiv preprint arXiv:2604.02286},
  year={2026}
}

@article{mcgregor2020mdine,
  title={MDiNE: a model to estimate differential co-occurrence networks in microbiome studies},
  author={McGregor, Kevin and Labbe, Aur{\'e}lie and Greenwood, Celia MT},
  journal={Bioinformatics},
  volume={36},
  number={6},
  pages={1840--1847},
  year={2020},
  publisher={Oxford University Press}
}

@article{espinoza2022differential,
  title={Differential network analysis of oral microbiome metatranscriptomes identifies community scale metabolic restructuring in dental caries},
  author={Espinoza, Josh L and Torralba, Manolito and Leong, Pamela and Saffery, Richard and Bockmann, Michelle and Kuelbs, Claire and Singh, Suren and Hughes, Toby and Craig, Jeffrey M and Nelson, Karen E and others},
  journal={PNAS nexus},
  volume={1},
  number={5},
  pages={pgac239},
  year={2022},
  publisher={Oxford University Press}
}

@article{article,
author = {Giuliani, Alessandro and Di Paola, Luisa and Setola, Roberto},
year = {2009},
month = {12},
pages = {235-245},
title = {Proteins as Networks: A Mesoscopic Approach Using Haemoglobin Molecule as Case Study},
volume = {6},
journal = {Current Proteomics},
doi = {10.2174/157016409789973743}
}

@String{Springer = "Springer-Verlag" }

@inproceedings{agapito2013cloud4snp,
  title={Cloud4SNP: distributed analysis of SNP microarray data on the cloud},
  author={Agapito, Giuseppe and Cannataro, Mario and Guzzi, Pietro Hiram and Marozzo, Fabrizio and Talia, Domenico and Trunfio, Paolo},
  booktitle={Proceedings of the International Conference on Bioinformatics, Computational Biology and Biomedical Informatics},
  pages={468--475},
  year={2013}
}

\end{document}